# Forced Measurement of Astronomical Sources at Low Signal-to-noise

A. Dutta , J. R. Peterson , and G. Sembroski
Department of Physics and Astronomy, Purdue University, West Lafayette, IN 47907, USA; dutta26@purdue.edu

## Abstract

We propose a modified moment matching algorithm to avoid catastrophic failures for sources with a low signal to noise ratio. The proposed modifications include a method to eliminate nonphysical negative pixel values and a forced single iteration with an initial guess derived from coadd measurements when iterative methods are unstable. We correct for all biases in measurements introduced by the method. We find that the proposed modifications allow the algorithm to avoid catastrophic failures in nearly 100% of the cases, especially at low SNR. Additionally, with a reasonable guess from coadd measurements, the algorithm measures the flux, centroid, size, shape, and ellipticity with a bias statistically consistent with zero. We show that the proposed method allows us to measure sources 7 times fainter than traditional methods when applied to images obtained from WIYN-ODI. We also present a scheme to find uncertainties in measurements when using the new method to measure astronomical sources.

*Unified Astronomy Thesaurus concepts:* Algorithms (1883); Astronomical methods (1043); Interdisciplinary astronomy (804)

## 1. Introduction

In optical astronomy, a series of short exposures are often taken to mitigate against changing observing conditions, avoid saturation, and study time variation. Stacking or coadding images has been used to measure extremely faint sources not detectable in individual images. A significant amount of work has been done on how to best coadd images (Fischer & Kochanski 1994; Hook & Lucy 1994; Bertin et al. 2002; Gruen et al. 2014; Zackay & Ofek 2017). However, coadds suffer from various drawbacks, such as averaging of systematics, loss of information, and introduction of artifacts. The presence of clouds, different seeing conditions, and different science requirements makes weighting of individual frames in coadds challenging (Zackay & Ofek 2017). This is especially a problem for large-scale surveys that use multiple visits, like the Sloan Digital Sky Survey (SDSS; Gunn & Weinberg 1994; York et al. 2000; Stoughton et al. 2002), PanSTARRS (Kaiser et al. 2002; Chambers et al. 2016), the Dark Energy Survey (DES; The Dark Energy Survey Collaboration 2005; Dark Energy Survey Collaboration et al. 2016), and the Rubin Observatory Legacy Survey of Space and Time (LSST Science Collaboration et al. 2009; Ivezić et al. 2019). Over the course of several years of operation, these surveys have produced or will produce hundreds to thousands of images of any specific region of the sky. Survey images are obtained under various conditions of seeing, clouds, air mass, and background noise. Traditionally, these images have been coadded to produce a final image on which all source detection and measurements are done (Annis et al. 2014; Waters et al. 2020) However, simply coadding images represents a significant loss of information. For instance, combining an image with a large seeing or ill-behaved point-spread function (PSF) with an image with particularly good seeing degrades the effective PSF significantly.

In theory, detecting the sources in the coadd and measuring them in individual exposures can improve measurements significantly. This is common for photometry where the zero-points are determined for each image individually to take into account the significant variation from one image to another (Stoughton et al. 2002; Skrutskie et al. 2006). Ivezić et al. (2007) were able to produce extremely accurate photometry from SDSS by taking into account systematic effects in individual images. Sako et al. (2008) used forced positional photometry on individual SDSS images to place upper limits on the magnitude of presupernova candidates.

Similarly, for astrometry, it has been shown that using multiple short exposures provides better astrometric estimates. In a long-exposure image, near-Earth asteroids appear as streaks. Inferring the average position of a source from a streak is inherently problematic and introduces large astrometric errors. The "shift-and-stack" method was first proposed by Tyson et al. (1992) and later used by Cochran et al. (1995) and Gladman et al. (1998) to detect Kuiper Belt and trans-Neptunian objects, respectively. More recently, Zhai et al. (2014) and Shao et al. (2014) used a similar technique called synthetic tracking, where each individual exposure is shifted and added such that the asteroids appears as point sources while stars appear as streaks in the final coadd. The asteroid position can then be measured with much better accuracy. To measure the relative position of the asteroid with respect to stars, one can average the position on stars in each individual exposure rather than measuring the streak stars in the coadded image and hence avoid measuring streaks altogether.

This approach of using multiple short exposures is also useful in gravitational lensing studies, where measuring shapes of astronomical sources is of paramount importance (Tyson et al. 2008; Jee & Tyson 2011; Mandelbaum 2018). This method also helps to take into account any chip-based systematic effects and discard/deweight periods of bad observing conditions. This idea has been used in CFHTLenS (Miller et al. 2013) and DES (Zuntz et al. 2018). Thus, in theory, all of the quantities you can measure (flux, position, shape) are enhanced by individual exposures.

The major challenge of using individual exposures is that the sources are much fainter than the coadd. Faint sources can be challenging to measure. Several methods have been proposed to measure faint sources (Hogg & Turner 1998; Lang et al. 2009;

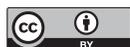

Teeninga et al. 2016; Lindroos et al. 2016). Some of the methods, such as forced photometry (Stoughton et al. 2002; Sako et al. 2008; Saglia et al. 2012; Lang et al. 2016), are simple but are limited to photometry. Other algorithms propose complicated methods to achieve the desired results (Hogg & Turner 1998; Teeninga et al. 2016).

In this paper, we propose a moment matching method that can measure the flux, centroid, size, and ellipticity of sources to an arbitrarily low signal to noise ratio (S/N). The paper is organized as follows. In Section 2, we explain the moment matching algorithm in detail. We also give an expression for errors in measurement when using this algorithm. In Section 3, we describe two novel modifications to use this algorithm for an arbitrarily low S/N. Finally, in Section 4, we show that our method works well with simulations in the presence of noise. Performance of this method on point sources in data obtained from WIYN-ODI is also shown.

## 2. Moment Matching Algorithm

In this section, we describe the moment matching algorithm used. This is similar to the one proposed by Kaiser et al. (1995) and later modified by Bernstein & Jarvis (2002). We use an adaptive moment matching method that uses an elliptical Gaussian as weight. The shape, size, and centroid of the elliptical Gaussian are iteratively calculated from the image itself. The choice of elliptical Gaussian profile is not unique and can be replaced with other popular profiles such as de Vaucouleurs and pure exponential. It has been shown that a weighted centroiding method similar to this has optimal performance when compared to other existing methods (Thomas et al. 2006). Several different method of centroiding can be found in the literature (King 1983; Stone 1989; Mighell 2005; Thomas et al. 2006).

We define $I(x,y)$ as the input image or 2D array, $\mu_x$ as the centroid position along the $x$-axis, $\mu_y$ as the centroid position along the $y$-axis, $\sigma_{xx}^2$ as the variance along the $x$-axis, $\sigma_{yy}^2$ as the variance along the $y$-axis, $\sigma_{xy}^2$ as the variance along the $y = x$ line, $w(x, y)$ as the weight value at the pixel $(x, y)$, and $B$ as the background level of the image. We also define $\mu_x = \mu_y = 0$ at the center of the image.

Initial guess values can be specified by the user. This is important for the forced measurement method described in the next section, since it relies on a reasonably good initial guess obtained from the coadd. Unless specified, the default values are $\sigma_{xx}^2 = 9$, $\sigma_{yy}^2 = 9$, $\sigma_{xy}^2 = 0$, $\mu_x = 0$, and $\mu_y = 0$. The background, $B$, is initially set to the median value of all pixels in $I(x, y)$. Next, the 2D elliptical Gaussian weight $w(x, y)$ is calculated using the values of $\sigma_{xx}^2$, $\sigma_{yy}^2$, $\sigma_{xy}^2$, $\mu_x$, and $\mu_y$,

$$
\begin{aligned}
w(x, y) = & \frac{1}{2\pi\sqrt{(1 - \gamma^2)\sigma_{xx}^2\sigma_{yy}^2}} \\
& \times \exp\Bigg(-\frac{1}{2[1 - \gamma^2]}\Bigg[\frac{(x - \mu_x)^2}{\sigma_{xx}^2} + \frac{(y - \mu_y)^2}{\sigma_{yy}^2} \\
& - 2\gamma\left(\frac{x - \mu_x}{\sigma_{xx}}\right)\left(\frac{y - \mu_y}{\sigma_{yy}}\right)\Bigg]\Bigg),
\end{aligned}
\tag{1}
$$

where

$$
\gamma = \frac{\sigma_{xy}^2}{\sigma_{xx}\sigma_{yy}}.
\tag{2}
$$

The following quantities are then calculated and updated iteratively:

$$
\mu_x = \frac{\sum_{x,y} x[I(x, y) - B]w(x, y)}{\sum_{x,y}[I(x, y) - B]w(x, y)},
\tag{3}
$$

$$
\mu_y = \frac{\sum_{x,y} y[I(x, y) - B]w(x, y)}{\sum_{x,y}[I(x, y) - B]w(x, y)},
\tag{4}
$$

$$
\sigma_{xx}^2 = 2\left(\frac{\sum_{x,y} xx[I(x, y) - B]w(x, y)}{\sum_{x,y}[I(x, y) - B]w(x, y)} - \mu_x^2\right),
\tag{5}
$$

$$
\sigma_{yy}^2 = 2\left(\frac{\sum_{x,y} yy[I(x, y) - B]w(x, y)}{\sum_{x,y}[I(x, y) - B]w(x, y)} - \mu_y^2\right),
\tag{6}
$$

$$
\sigma_{xy}^2 = 2\left(\frac{\sum_{x,y} xy[I(x, y) - B]w(x, y)}{\sum_{x,y}[I(x, y) - B]w(x, y)} - \mu_x\mu_y\right),
\tag{7}
$$

$$
N = \frac{\sum_{x,y}[I(x, y) - B]w(x, y)}{\sum_{x,y} w^2(x, y)},
\tag{8}
$$

$$
B = \frac{\sum_{x,y} I(x, y) - N}{\sum_{x,y} 1}.
\tag{9}
$$

We repeat this until the values of both $\sigma_{xx}^2$ and $\sigma_{yy}^2$ change by $<10^{-3}$. This is defined as convergence.

It is important to quantify the errors in measurement of flux, centroid, size, and ellipticity when using this algorithm. We do this in a semiempirical way using simulations. We simulate a large number of sources with various background, flux, and ellipticity and measure the uncertainty in these as a function of flux, size, background, size, and ellipticity. The range of flux is 1–10⁷ counts, the size is 10–20 pixels, and ellipticity is drawn from a Gaussian distribution centered at 0.1 with a width of 0.2. Ellipticity values of less than 0 are rejected.

Simulating sources is done in two steps. First we simulate the background, and then the object is added. To simulate the background, each pixel is assigned a random real number drawn from a Gaussian distribution of mean $B$ and variance $B$. To simulate an object of flux $N$, a random number is drawn from a Gaussian distribution with mean $N$ and variance $N$. Say $N_1$ is the total number of photons received from the object. Next, a probability distribution function is created with the same 2D shape as the object, and $N_1$ photons are drawn from this distribution to simulate the source. We use the multivariate normal distribution method in numpy (Harris et al. 2020) to perform this. Next, we run the abovementioned moment matching algorithm on the source for a single iteration with the correct guess shape. This allows us to isolate the errors in various measured parameters, even when one has almost perfect knowledge of the source.

The error when measuring image parameters arises mainly from photon noise and background. Hence, we simulate a million sources with no background to recover the error due to photon noise. To recover the background component, we simulate sources with background, find the error, and subtract the error due to photon noise. An alternate way to find the second component is to simulate sources without photon noise but with background and find the errors. Both methods give consistent results.

Dimensional analysis is used as a sanity check. For the image parameters being measured, flux is a dimensionless quantity denoted by $N$. Background is in units or $1/\text{pixel}^2$ and is denoted by $B$. The variance, namely, $\sigma^2_{xx}$, $\sigma^2_{yy}$, and $\sigma^2_{xy}$, has units of $\text{pixel}^2$. Also, we define $A = \pi\sqrt{\sigma^2_{xx}\sigma^2_{yy} - \sigma^4_{xy}}$ as the area of the object and $S = \sqrt{\sigma^2_{xx} + \sigma^2_{yy}}$ as the size of the object.

### 2.1. Flux

There are two significant sources of error in flux measurements: (1) a Poisson component due to source photons and (2) a Poisson component due to background photons. It is well known that the noise in flux due to photon noise is $\sqrt{N}$, where $N$ is the number of detector counts. However, the algorithm itself can only expect to asymptotically achieve this in the presence of noise from the background. Next, using the abovementioned method, we find that the error due to background is $\sqrt{4\pi S^2 B}$, where $S$ denotes size and $B$ denotes background. This can derived considering the expression for $N$ in Section 2:

$$N = \frac{\sum_{x,y}(I(x, y) - B)w(x, y)}{\sum_{x,y}w^2(x, y)}. \quad (10)$$

It can be shown that $\sum_{x,y}w^2(x, y) = 1/(4\pi S^2)$. We can write the previous equation as

$$N = 4\pi S^2 \sum_{x,y}(I(x, y) - B)w(x, y). \quad (11)$$

Assuming the uncertainty arises only due to background photons, the uncertainty in $N$ then is

$$\epsilon^2_N = (4\pi S^2)^2 \sum_{x,y} w^2(x, y)\sigma^2(x, y), \quad (12)$$

where $\sigma^2(x, y)$ is the uncertainty in counts of pixels $(x, y)$. Hence, $\sigma^2(x, y) = B$ for all pixels, and this can be further simplified to

$$\epsilon^2_N = (4\pi S^2)^2 B \sum_{x,y} w^2(x, y). \quad (13)$$

After simplification, the final expression for error due to background is

$$\epsilon^2_N = 4\pi S^2 B. \quad (14)$$

This expression is identical to the $\sqrt{4\pi b\sigma^2}$ expression obtained by Bernstein & Jarvis (2002), where $b$ is the background counts per pixel and $\sigma$ is the size of the source. The background Poisson component depends on the weight function. With our choice, the error is $\sqrt{4\pi S^2 B}$. This appears to be optimal. Hence, we introduce a constant $K$:

$$K = \frac{4\pi S^2 B}{N}. \quad (15)$$

The error in flux can be written as

$$\epsilon_{\text{flux}} = \sqrt{N(1 + K)}. \quad (16)$$

Using this, we can define S/N as

$$\text{S/N} = \frac{N}{\sqrt{N(1 + K)}}. \quad (17)$$

In Figure 1, we show that this expression accurately predicts errors in flux. The black circle that denotes $\epsilon^2_{\text{flux}}/N^2$ closely follows the $1/N$ line, second from the top. Deviations seen in Figure 1 are primarily due to the algorithmic limitation of not being able to characterize the background perfectly.

### 2.2. Size

The errors in size have two components as well, namely, intrinsic and background. The uncertainty in size can be written as

$$\epsilon_{\text{size}} = \sqrt{\frac{1}{2}\frac{S^2}{N}(1 + K)}, \quad (18)$$

where $N$ is the flux, $S$ denotes the size of the source, $B$ is the background, and $K$ is defined in Equation (15). The first component is due to the intrinsic uncertainty arising from photon noise, and the second component is due to the background. The reason we use size ($S$) over the conventional $\sigma$, i.e., spread along each axis, is that using size is more accurate than using $\sigma$ in cases of elliptical objects.

Shown in Figure 1, we find that the light green circles, which denote $\epsilon^2_{\text{size}}/S^4$, closely follow the $1/2N$ line, third from the top.

We also write down the formula for errors in sigma parameters as

$$\epsilon_{\sigma^2_{xx}} = \sqrt{\frac{S^4}{N}(1 + K)}, \quad (19)$$

$$\epsilon_{\sigma^2_{yy}} = \sqrt{\frac{S^4}{N}(1 + K)}, \quad (20)$$

$$\epsilon_{\sigma^2_{xy}} = \sqrt{\frac{1}{2}\frac{S^4}{N}(1 + K)}, \quad (21)$$

where $N$ is the number of counts, $B$ is the background, and $S$ is the effective size. The dark blue circle and dark blue triangle in Figure 1 denote $\epsilon^2_{\sigma^2_{xx}}/S^4$ and $\epsilon^2_{\sigma^2_{yy}}/S^4$, respectively, and follow the $1/N$ black line, second from the top. Similarly, $\epsilon^2_{\sigma^2_{xy}}/S^4$, shown in dark green circles, follows the $1/2N$ black line, third from the top.

### 2.3. Centroid

Once again, the errors in centroid arise from background and photon noise. According to Thomas (2004), the centroid uncertainty due to photon noise is $\sqrt{A/(\pi N)}$, where $A$ is the area and $N$ is the number of counts. We did recover this from our experiments as well. Extending this idea to include background, we have

$$\epsilon_{\mu_x} = \sqrt{\frac{S^2}{N}(1 + K)}, \quad (22)$$

where $N$ is the flux, $S$ denotes the size of the source, and $B$ is the background. This is also equivalent to the errors presented by King (1983) and Lang et al. (2009). We decide not to use $\sigma$, i.e., the spread along each axis, in this formula and instead use size ($S$) for the same reason as before. In Figure 1, $\epsilon^2_{\mu_x}/S^2$ is shown in light blue circles and seems to follow the $1/N$ line, second from the top. The $y$-component is shown in light blue triangles and follows the same line.

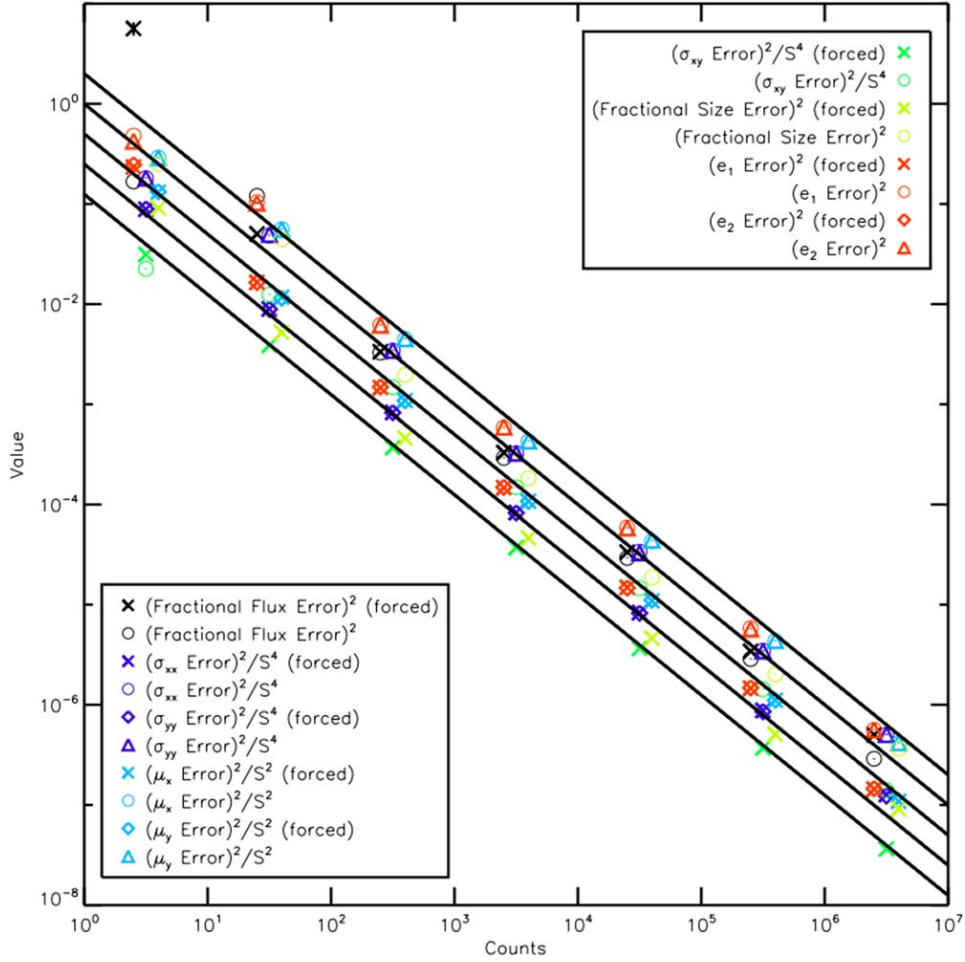

**Figure 1.** This plot shows the intrinsic errors in flux, centroid, size, and ellipticity. A wide range of sizes, counts, and ellipticity has been used to make this plot. The background was set to 0. To produce each point, $10^6$ samples were drawn, and the error in measurement was recorded. Forced indicates only a single iteration was performed with the true values as guess. The black lines from the top are $2/N$, $1/N$, $1/2N$, $1/4N$, and $1/8N$, respectively. The black circle that denotes $\epsilon^2_{flux}/N^2$ closely follows the $1/N$ line. Light green circles that denote $\epsilon^2_{size}/S^4$ closely follow the $1/2N$ line. In the case of the centroid, $\epsilon^2_{\mu_x}/S^2$ is shown in light blue circles and seems to follow the $1/N$ line. For ellipticity, $\epsilon^2_{e_1}$ is shown in red circles, and $\epsilon^2_{e_2}$ is shown in red triangles. Both follow the $1/N$ line. It is clear from the plot that the errors in all components except flux are approximately twice as large when convergence is allowed. The large deviation in fractional flux error when the number of counts is extremely low is likely due to the breakdown of the Gaussian approximation.

### 2.4. Ellipticity

The error in ellipticity is

$$\epsilon_{\text{ellipticity}} = \sqrt{\frac{2}{N}(1 + K)}.$$  (23)

In Figure 1, $\epsilon^2_{e_1}$ is shown in red circles, and $\epsilon^2_{e_2}$ is shown in red triangles. Both follow the $1/N$ black line (second from the top). The two components of ellipticity are $e_1$ and $e_2$ and are defined as

$$e_1 = \frac{\sigma^2_{xx} - \sigma^2_{yy}}{\sigma^2_{xx} + \sigma^2_{yy}},$$  (24)

$$e_2 = \frac{2\sigma^2_{xy}}{\sigma^2_{xx} + \sigma^2_{yy}}.$$  (25)

Total ellipticity is obtained by adding $e_1$ and $e_2$ in quadrature.

The error quoted above is when the algorithm is allowed to converge. We once again emphasize that the deviation of points in Figure 1 from the exact expression is primarily due to the limitation of our algorithm. We found that when we supplied the algorithm with the true values as the initial guess and ran for one iteration, the errors would decrease significantly. The form of the errors would remain the same, but the errors are reduced by a factor of 2. The only exception to this rule is the error in flux, where the errors remain unchanged. We also found a slight dependence of errors on ellipticity in some cases. For example, it is not hard to see when $e_1 > 0$; the error in $\mu_x$ will be larger than the error in $\mu_y$. These errors are significant at $e > 0.5$. Only a small fraction of astronomical sources have such high ellipticity.

## 3. Forced Measurement

### 3.1. Generalization of Forced Photometry to All Measurements

We found that the above algorithm fails at arbitrarily low S/N. This is expected but completely limits the ability to make individual frame measurements. The failure starts as rare events at S/N $\lesssim$ 10 and climbs up to a 100% failure rate at S/N $\leqslant$ 6 as shown in Figure 7. This poses a huge problem when trying to make measurements on individual frames, since a vast majority of the sources are expected to be extremely faint in individual frames. To solve this, we generalize a technique originally developed for forced photometry (Stoughton et al. 2002; Lupton et al. 2005; Sako et al. 2008; Saglia et al. 2012; Lang et al. 2016). We call this method forced measurement to generalize it to photometry, astrometry, and size/shape measurements.

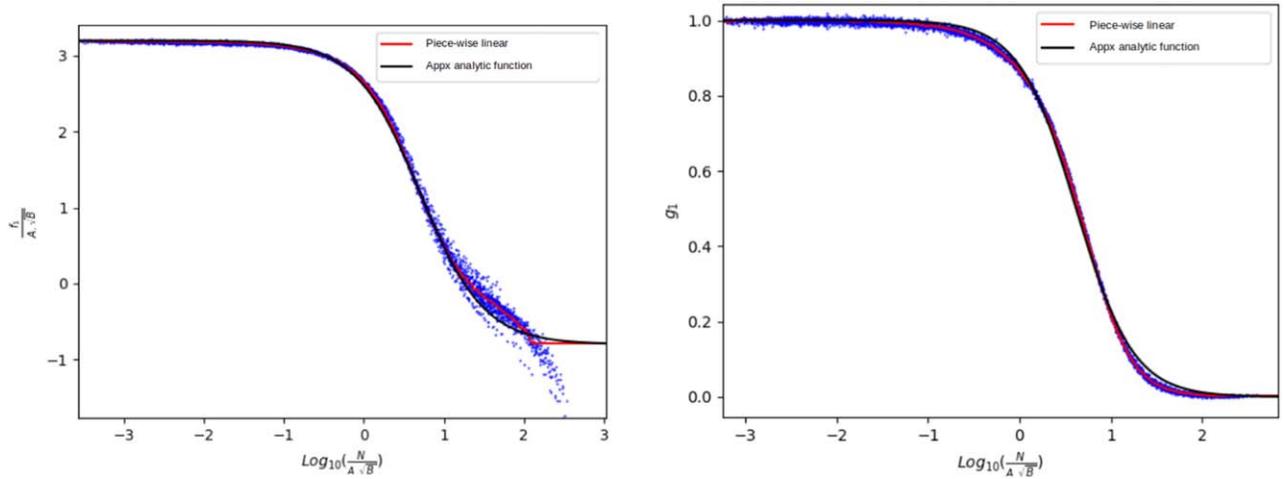

**Figure 2.** The figure on the left shows how the flux overestimation function $f_1$ is related to other image parameters. On the right, we show how the $\sigma^2$ overestimation function $g_1$ is related to other image parameters. The points in blue show the 4512 different cases we simulated to find the correction. The red curve is the piecewise linear interpolation. The black line shows the approximate analytical function in Equations (31) and (32).

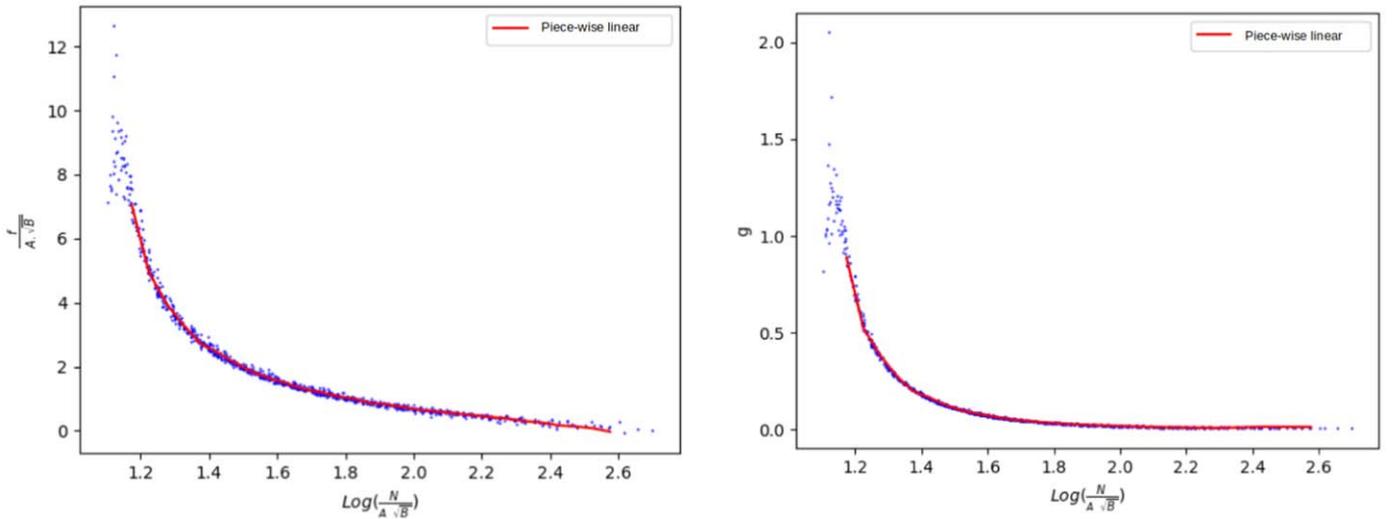

**Figure 3.** The correction factor $f_\infty$ when convergence is allowed is shown on the left. The correction factor $g_\infty$ when convergence is allowed is shown on the right. Clearly, both factors blow up at low S/Ns, implying that convergence is only feasible at a fairly high S/N.

Forced photometry is a method of measuring flux when reliable measurements cannot be made. Forced photometry uses the position of the sources detected, often in coadd or other data, to then guide where flux measurements can be made in individual frames. This can be effective even when sources cannot be detected in individual frames. There are several different types of photometry, such as aperture photometry (Stetson 1987), PSF photometry, and model photometry (Lupton et al. 2001; Skrutskie et al. 2006; Ivezić et al. 2007).

Forced photometry can use any of these slightly different methods. Since we use an elliptical Gaussian weight for moment matching, we concentrate on model photometry, which also uses an elliptical Gaussian model. Such a model photometry would be equivalent to running the moment matching algorithm for one iteration.

Using the motivation of forced photometry, we note that we can generalize this approach for astrometry and shape measurements. As noted before, if forced photometry, which is equivalent to running the moment matching algorithm for one iteration, produces sensible flux measurements, we expect the other measurements to be sensible as well, albeit noisy. In other words, at extremely low S/N, where the traditional moment matching algorithm fails to converge, we can expect to measure the position and shape of a source by running the moment matching algorithm for one iteration. As is the case with forced photometry, we need the position and shape of the source to perform forced measurement. The position of the source on the sky and its shape (profile) is determined from the coadd or perhaps some other deeper image. We assume that in the coadded image the S/N is sufficiently high for the traditional moment matching algorithm to converge. This allows us to obtain a good estimate/guess of the flux, shape, and centroid in each individual image of the coadd. In individual frames, we create a cutout centered on this position and use the coadd shape as a guess (taking into account the necessary PSF correction) to run the algorithm for one iteration. The flux obtained using this strategy is identical to using model forced photometry. In addition to flux, this strategy was found to yield fairly accurate values of shape and centroid using the equations for the moment matching algorithm in Section 2, as long as our cutouts are fairly centered and the input profile is not too far from the true profile. In the following discussions, we use input profile and guess shape interchangeably.

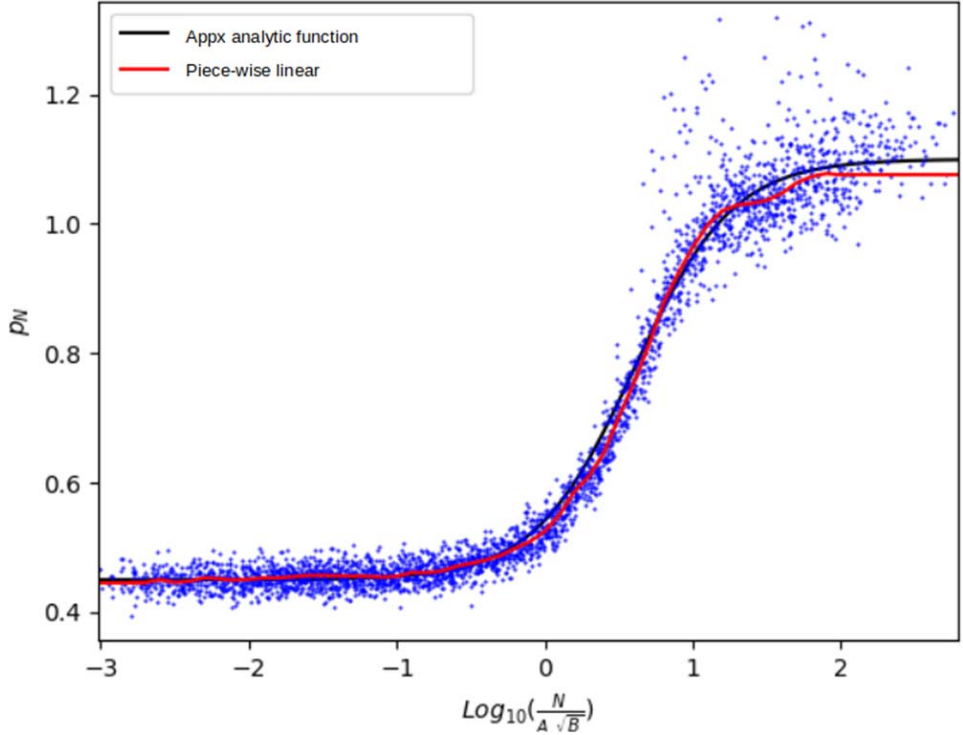

**Figure 4.** Error adjustment factor for flux, i.e., $p_N$, is shown. This is the factor by which the error in flux presented in Section 2.1 needs to be adjusted. The adjustment factor is needed because the initial guess is the truth, and a single iteration is performed when using forced measurement. The red curve is the piecewise linear interpolation. The approximate logistic function is shown in Equation (36).

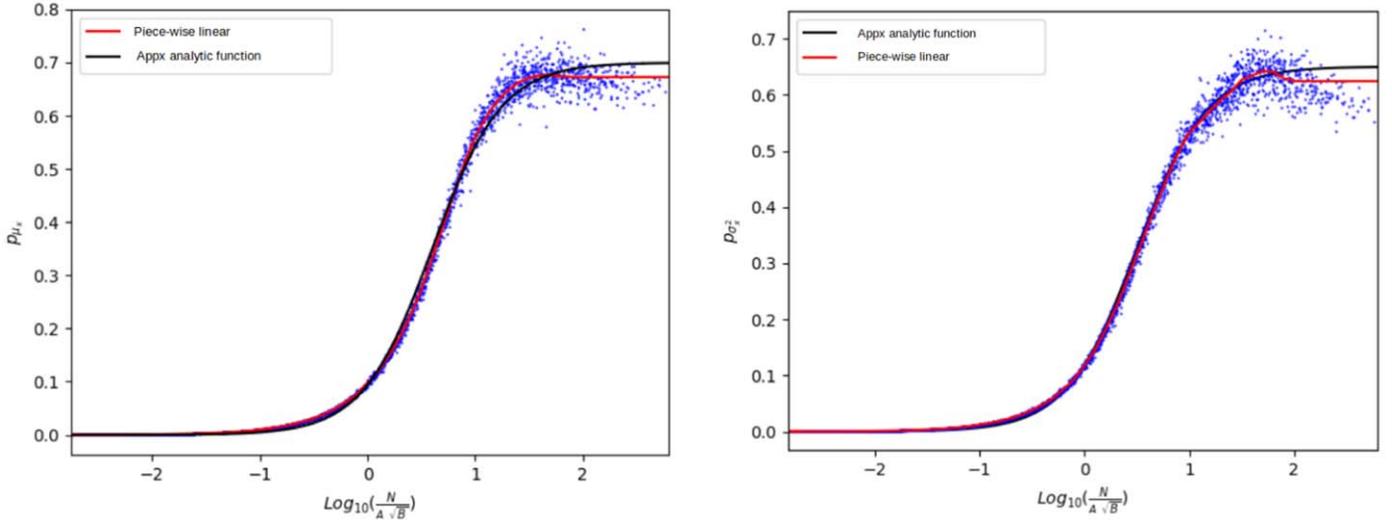

**Figure 5.** Error adjustment factor for the centroid, i.e., $p_{\mu_i}$, on the left and sigmas, i.e., $p_{\sigma_{xx}^2}$, on the right. The two are basically identical but are shown separately for the sake of clarity. The red curve is the piecewise linear interpolation. The approximate logistic function is shown in Equations (37) and (38).

### 3.2. Truncation Correction

While the above strategy helps us push the limits, even this starts failing at S/N $\lesssim$ 5. At such a low S/N, by pure chance, $\sigma_{xx}^2$ and $\sigma_{yy}^2$ yield negative values sometimes. Such a result has no physical significance, and the problem gets worse as the S/N is decreased. We find that pixels with negative residual values after background subtraction cause this issue. It is clear that negative pixel values, after accurate background subtraction, have no physical meaning and arise from Poisson fluctuation.

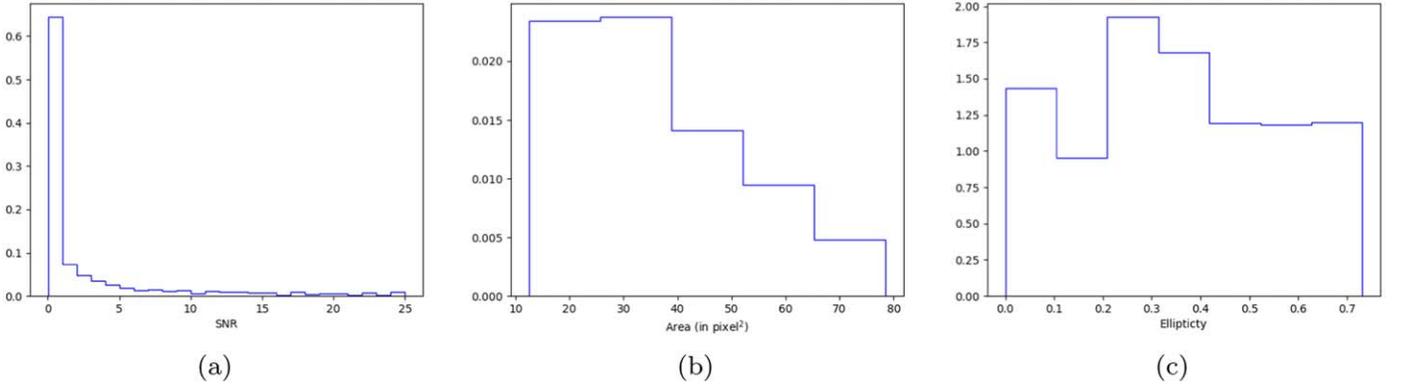

**Figure 6.** (a) The normalized S/N histogram for all 12,773 sources is shown. These sources are used to gauge the performance of the traditional moment matching algorithm, the forced measurement algorithm run until convergence, and the forced measurement algorithm. This graph has been truncated at S/N = 25. Similarly, (b) and (c) show the normalized histograms of the area and ellipticity of the sources, respectively.

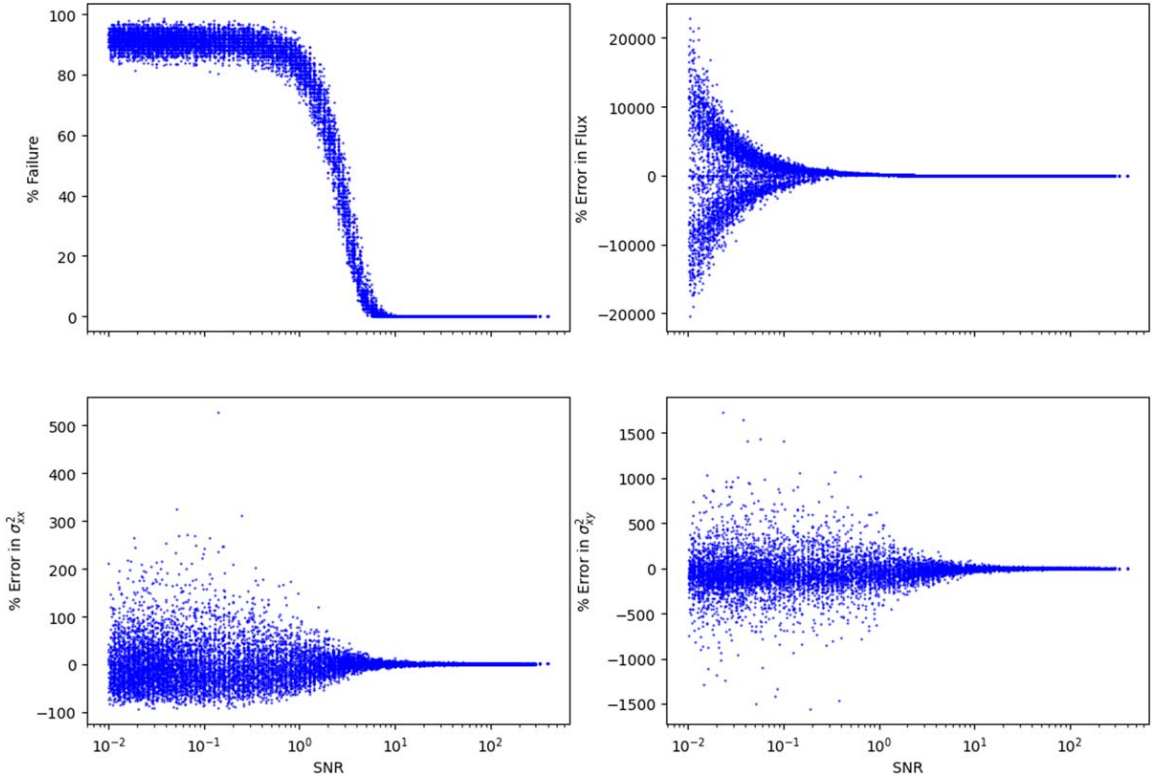

**Figure 7.** In these plots, we show the performance of the traditional moment matching algorithm for 12,773 cases. We vary the size from 2 to 5 pixels and flux from 1 to $10^5$ counts. Background varies from 50 to 2000 counts. Ellipticity values vary from 0 to 0.7 distributed between both $e_1$ and $e_2$. The upper left plot shows the failure percentage as a function of S/N. The upper right plot shows the percent deviation of the flux from the true value. The lower left and lower right plots show the percentage error in $\sigma_{xx}^2$ and $\sigma_{xy}^2$ as a function of S/N, respectively. It is clear that at S/N $\lesssim$ 10, catastrophic failures become more common, and the errors also become very large, as expected, sometimes even exceeding 100%. In the range $0.1 < $ S/N $ < 1$, the failure rate is $89.5\% \pm 3.6\%$, and the percentage error in flux is $396\% \pm 474\%$. The error in determining $\sigma_{xx}^2$ is $-1.9\% \pm 48\%$, while the percentage error for $\sigma_{xy}^2$ is $-40\% \pm 211\%$.

Consider a single pixel having $N$ counts, of which $B$ counts are from background. Background-subtracted counts are hence $(N - B) \pm \sqrt{N}$. To ensure we never end up with negative pixel values, which are unphysical, we take only the positive portion of this distribution and call the median value the true pixel value. When $N - B >> 0$, the median value is basically the same as $N - B$. However, as $N - B$ becomes closer to 0 and then negative, the closer to 0 the median value gets. Solving this give us

$$\text{erf}\left(\frac{N - B - \mu}{\sqrt{2N}}\right) = \frac{1}{2}\,\text{erf}\left(\frac{N - B}{\sqrt{2N}}\right) - \frac{1}{2}, \qquad (26)$$

where $N$ is the pixel value, $B$ is the background, $\mu$ is the inferred pixel value, and erf signifies the error function. Our objective is to calculate $\mu$ given $N$ and $B$. This can be simplified

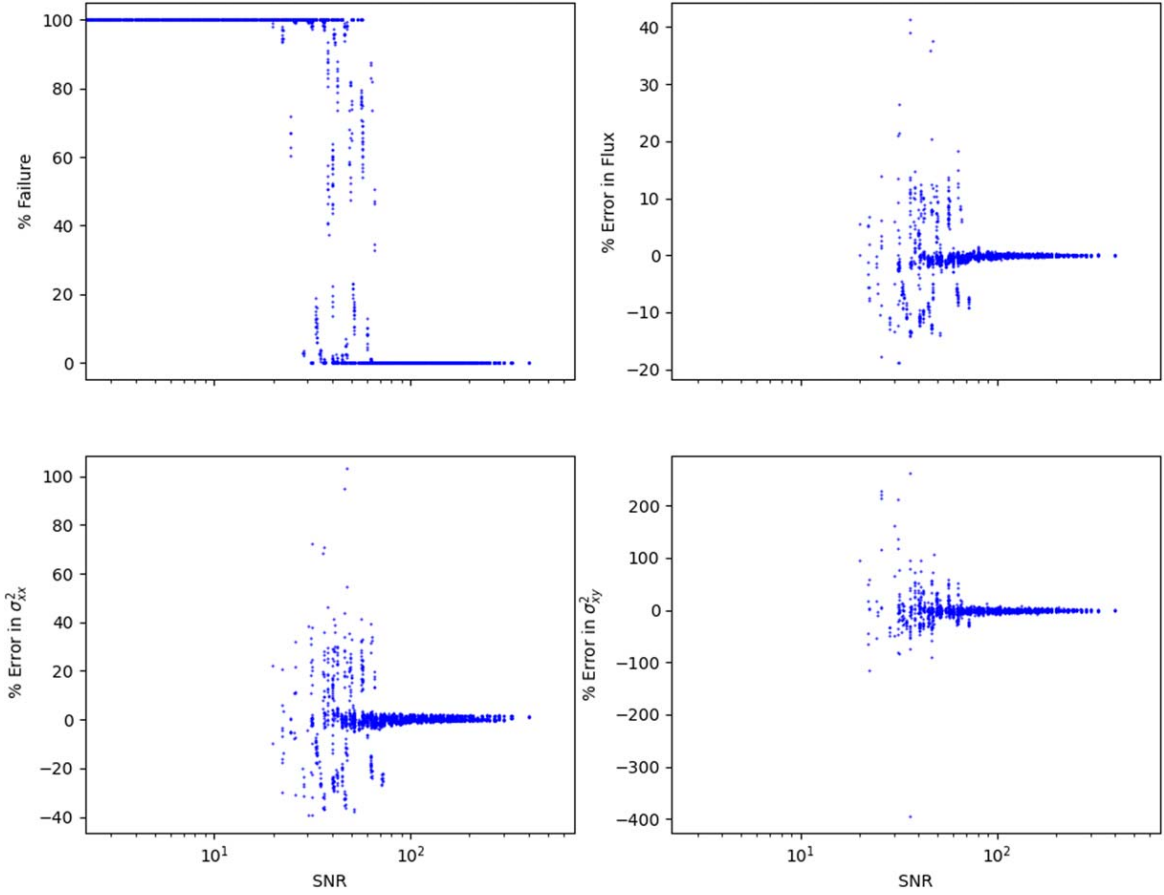

**Figure 8.** In these plots, we show the performance of our modified moment matching algorithm run until convergence for the same range of cases as Figure 7. We find that convergence is consistent only when S/N $\gtrsim$ 100. This shows that forced measurement does not perform well when applied iteratively, unless the source is at extremely high S/N $\gtrsim$ 100.

to

$$\mu = N - B - \sqrt{2N} \left\{ \text{erf}^{-1} \left[ \frac{1}{2} \, \text{erf} \left( \frac{N-B}{\sqrt{2N}} \right) - \frac{1}{2} \right] \right\}. \quad (27)$$

However, this equation is difficult to calculate and is inherently unstable due to the inverse error function. Hence, we fit a function to this, which is shown as

$$\mu = \begin{cases} N - B + \sqrt{2N} \, \dfrac{0.477}{e^{0.928q + 0.247q^2 + 0.04q^3}} & \text{when } (N-B) > 0 \\[2mm] \sqrt{2N} \, \dfrac{0.477}{e^{0.551q - 0.06q^2 + 0.003q^3}} & \text{when } (N-B) \leqslant 0 \end{cases}, \quad (28)$$

where $q = (N - B)/\sqrt{N}$. Using Equation (28) ensures that we always have positive pixel values and hence avoids the problem of negative $\sigma_{xx}^2$ or $\sigma_{yy}^2$. By forcing all pixels to have positive values, the size and flux are systematically overestimated. It was found that the overestimation was a function of flux, the sigmas of the source, and the background. We quantify the overestimation as follows:

$$N_t = N_m - f_n(N_t, A_t, \sqrt{B}), \quad (29)$$

$$\sigma_t^2 = \sigma_m^2 - g_n(N_t, A_t, \sqrt{B})\sigma_t^2, \quad (30)$$

where $\sigma^2$ can be replaced with $\sigma_{xx}^2$, $\sigma_{yy}^2$, or $\sigma_{xy}^2$ and $n$ is the number of iterations. $N_t$ refers to the true number of photons received from the source, and $A_t$ refers to the true area of the source. $N_m$ is the measured value of $N$ after performing one iteration of the moment matching algorithm on the truncated image; similarly, $\sigma_m^2$ is the measured shape parameter. $B$ stands for the median background value. The actual form is more subtle, and it was found that $f_n/(A_t\sqrt{B})$ is a function of log $(N/(A_t\sqrt{B}))$, whereas $g_n$ is a function of log $(N/(A_t\sqrt{B}))$. This is shown in Figure 2 for $n = 1$. In practice, knowing $N_t$ and $A_t$ beforehand is impossible. It was found that using the guess values derived from the coadd works well. These are the same initial guess values we use for forced measurement.

The functions $f$ and $g$ are dependent on the number of iterations. This can be attributed to the fact that we forced all the pixels to have positive values using Equation (27); hence, the $\sigma$ values used at the start of a new iteration keep increasing. This causes a runaway effect for most cases except when S/N $\gtrsim$ 100. A similar effect is seen for most other parameters, such as background and flux. It was found that when the S/N is higher than ~100, the algorithm converges in less than 100 iterations. The $f$ and $g$ functions for the one iteration case are denoted by $f_1$ and $g_1$. When convergence is allowed, the $f$ and $g$ functions are denoted as $f_\infty$ and $g_\infty$, respectively.

The graphs for $f_1$ and $g_1$ are shown in Figure 2. In order to produce these graphs, we simulated 4512 sources with values of flux ranging from 1 to $10^5$ counts, background values ranging from 50 to 2000 counts, size ranging from 2 to 6 pixels, and ellipticity ranging from 0 to 0.5. Each case was run

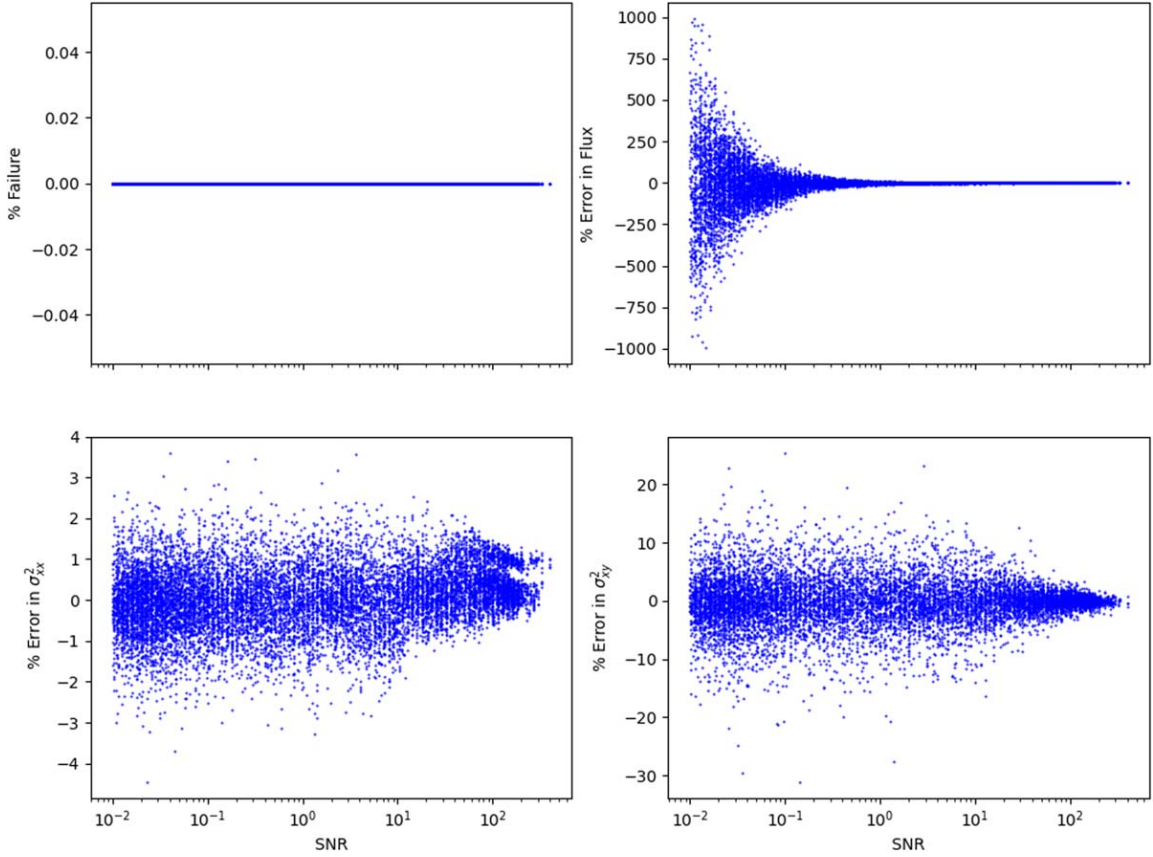

**Figure 9.** In these plots, we show the performance of the forced measurement algorithm for the same range of cases as Figure 7. Even at an extremely low S/N, unlike the traditional moment matching algorithm, we manage to avoid catastrophic failures as shown in the first graph. This is crucial for measuring ultrafaint sources. The error does increase significantly at low S/N. In the range $0.1 < S/N < 1$, the failure rate is 0%, and the percentage error in flux is $-1.9\% \pm 21\%$. In the same range, the error in $\sigma_{xx}^2$ is $0\% \pm 0.8\%$, and the error in $\sigma_{xy}^2$ is $-0.2\% \pm 4.7\%$.

250 times, and we measured the median overestimation of flux and $\sigma$. From that, we calculate $f$ and $g$ as mentioned in Equations (29) and (30). To make the piecewise linear curve in Figure 2, we divide the x-axis into sections of 0.05 and get the median of all points in that section. We join all such points to obtain the red curve in Figure 2. To produce the black lines, we fit a logistic function shown in Equations (31) and (32):

$$f_1 = \frac{4}{1 + e^{2.5(x-0.7)}} - 0.8, \qquad (31)$$

$$g_1 = \frac{1}{1 + e^{3.5(x-0.62)}}, \qquad (32)$$

where $x = \log_{10}[N/(A\sqrt{B})]$.

The graphs for $f_\infty$ and $g_\infty$ are shown in Figure 3. We follow the same process as before, except we now let the algorithm run until convergence. The condition for convergence was defined as when both $\sigma_{xx}^2$ and $\sigma_{yy}^2$ change by $<10^{-3}$ from one iteration to the next. It is clear from Figure 3 that the factors $f_\infty$ and $g_\infty$ become unstable and reach very high values on the left edge of the graph (low S/N).

### 3.3. Error Correction for Truncation

In Section 2, we presented the errors in various measured values when using a traditional moment matching algorithm until convergence. It was found that the error when using a single iteration of the forced measurement was much smaller. This partly comes from the fact that to run a single iteration of forced measurement, we need to input reasonably good initial guess values. It was found that the errors decrease by a factor, once again as a function of $f_n/(A_t . \sqrt{B})$. We decide to quantify the errors as follows:

$$\sigma_N = p_N \sqrt{N(1 + K)}, \qquad (33)$$

$$\sigma_{\sigma_{xx}^2} = p_{\sigma_{xx}^2} \sqrt{\frac{S^4}{N}(1 + K)}, \qquad (34)$$

$$\sigma_{\mu_x} = p_{\mu_x} \sqrt{\frac{S^2}{N}(1 + K)}, \qquad (35)$$

where $p_N$, $p_{\sigma_{xx}^2}$, and $p_{\mu_x}$ are the error adjustment factors for flux, $\sigma_{xx}^2$, and $\mu_x$, respectively. Once again, we follow the same procedure used to find functions as $f_1$ and $g_1$ in order to determine the three different $p$ functions. Figure 4 shows $p_N$, while Figure 5 shows $p_{\sigma_{xx}^2}$ and $p_{\mu_x}$ on the left and right, respectively. By using symmetry arguments, in the above equations, we can replace $\mu_x$ with $\mu_y$, and $\sigma_{xx}^2$ can be replaced with $\sigma_{yy}^2$. To get the error equation for $\sigma_{xy}^2$, we multiply the right-hand side of Equation (34) with $\sqrt{1/2}$. The approximate logistic forms for $p_N$, $p_{\sigma_{xx}^2}$, and $p_{\mu_x}$ were found to be

$$p_N = \frac{0.65}{1 + e^{-3(x-0.6)}} + 0.45, \qquad (36)$$

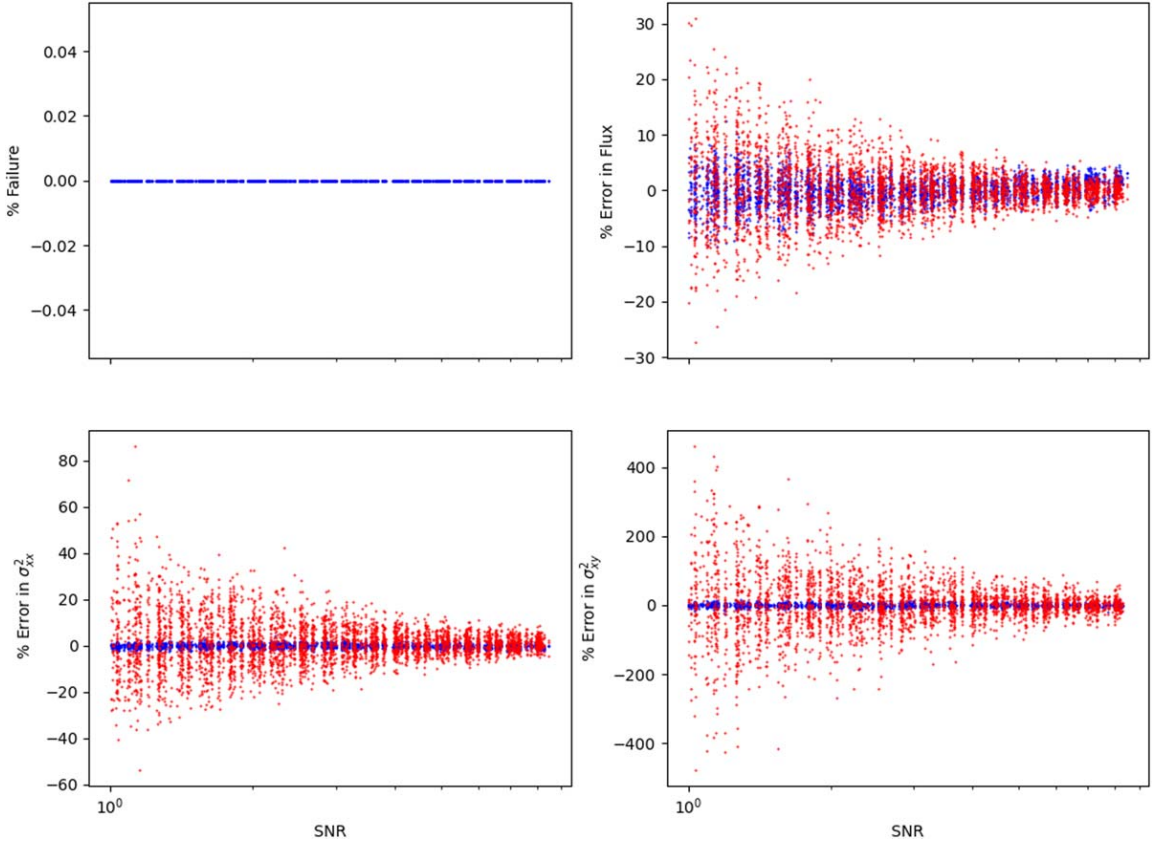

**Figure 10.** The red points show the coadd measurements, whereas blue points show the measurements obtained from the individual images. The x-axis is the S/N of individual images. When using forced measurement, we do not find any bias. The graphs clearly show the better error properties obtained when using this method.

$$p_{\sigma_{xx}^2} = \frac{0.65}{1 + e^{-3(x-0.5)}}, \tag{37}$$

$$p_{\mu_x} = \frac{0.7}{1 + e^{-3.1(x-0.6)}}, \tag{38}$$

where $x = \log_{10}[N/(A\sqrt{B})]$.

## 4. Validation

### 4.1. Simple Tests

In this section, we show the characteristics of the three different algorithms under different values of S/N: (1) the traditional moment matching algorithm, (2) the forced measurement algorithm when it is run until convergence, and (3) the forced measurement algorithm, i.e., with a single iteration. For this test, we simulate 12,773 sources with varying size, flux, ellipticity, and background. We vary the size from 2 to 5 pixels and flux from 1 to $10^5$ counts. Background varies from 50 to 2000 counts. Ellipticity values vary from 0 to 0.7 distributed between both $e_1$ and $e_2$. The distributions of S/N, area, and ellipticity of the sources are shown in Figure 6. For each source, we produce 200 instances and apply the three different algorithms. For the traditional moment matching algorithm and forced measurement until convergence, we start with the guess values mentioned in Section 2. For the forced measurement single iteration algorithm, the true values were supplied as a guess.

We measure four key parameters the failure rate and the percentage deviation of flux, $\sigma_{xx}^2$ and $\sigma_{xy}^2$ from the truth. We define failure as when the algorithm produces nonsense results, such as negative values of $\sigma_{xx}^2$ or $\sigma_{yy}^2$, or the algorithm crashes. If the $\sigma_x$ or $\sigma_y$ value is larger than one-third of the cutout size, we also classify that as failure. The cutout size is $100 \times 100$ pixels for all cases. The results are shown in Figures 7, 8, and 9. We see in Figure 7, with the traditional moment matching algorithm, that the failure rate begins to climb at S/N $\sim 10$. In the range $10 < $ S/N $ < 15$, the failure rate is 0%, while the percentage error in flux is $1\% \pm 1.1\%$. In the same range, the error in $\sigma_{xx}^2$ is $0.9\% \pm 2.3\%$, while the percentage error in $\sigma_{xy}^2$ is $-2.4\% \pm 15\%$. At lower S/N, the error in flux begins to climb rapidly and approaches nonsense values of $>10^4\%$. Errors in $\sigma_{xx}^2$ and $\sigma_{xy}^2$ show the same behavior. In the range $0.1 < $ S/N $ < 1$, the failure rate is $89.5\% \pm 3.6\%$, and the percentage error in flux is $396\% \pm 474\%$. In the same range, the percentage error in $\sigma_{xx}^2$ is $-1.9\% \pm 48\%$, while the percentage error in $\sigma_{xy}^2$ is $-40\% \pm 211\%$.

Forced measurement run until convergence is shown in Figure 8. We find that this method is able to produce accurate results only at high S/N $\gtrsim 100$. This algorithm is unstable because repeated iterations along with distortion correction cause the size to become progressively larger or smaller until it exceeds the effective size of the cutout or becomes smaller than the size of a single pixel. At higher S/N, we get only a small error in flux and $\sigma$. We explored the stability of this method at lower S/N and found that using the truncation once and slightly overestimating the background would make the failure rate comparable to the traditional moment matching algorithm. However, the bias correction arising from truncation cannot be obtained unless the true values are known.

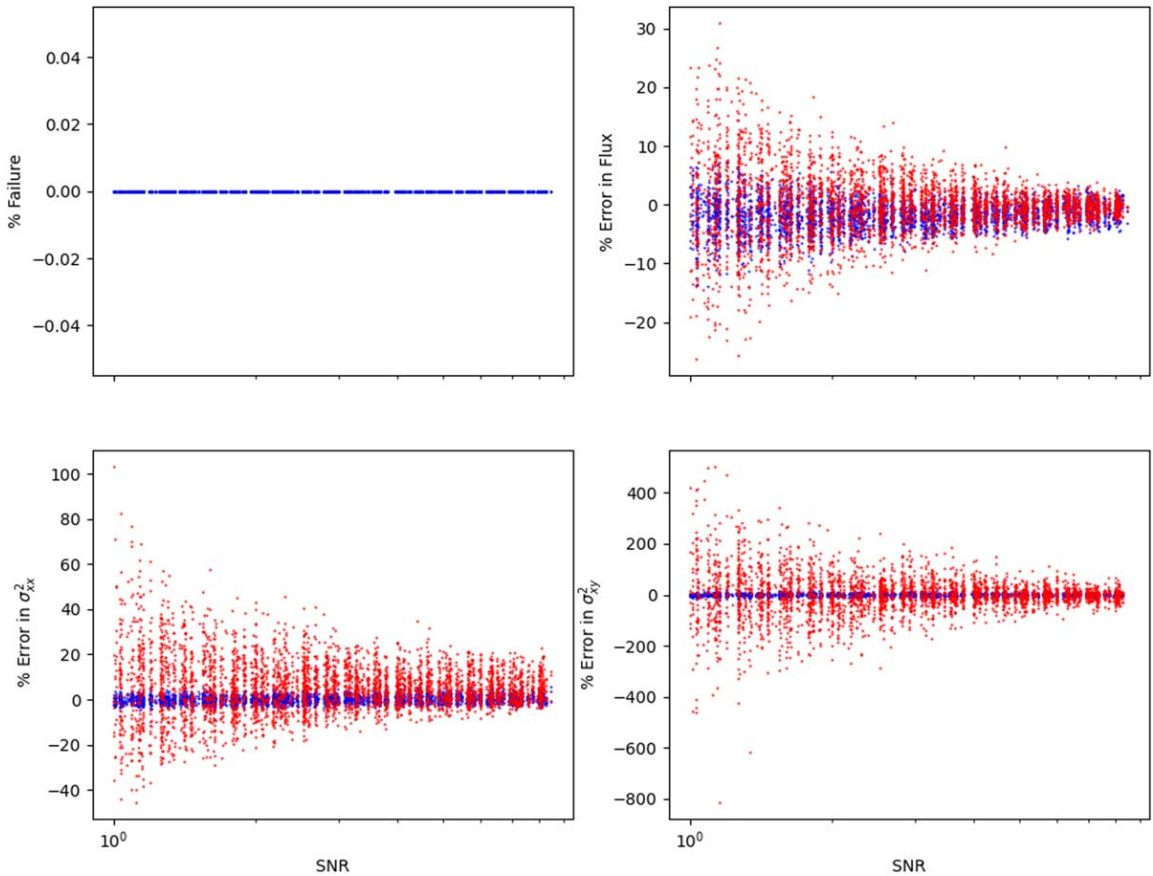

**Figure 11.** The red points show the coadd measurements, whereas blue points show the measurements obtained from the individual images. The *x*-axis is the S/N of individual images. We also inject an error of 15% in guess flux and $\sigma^2$ parameters. When simulating the individual sources, an error of 15% is injected in flux and $\sigma$, and an error of 1 pixel introduced in the centroid. We see a slight bias of $-2.9\% \pm 4.1\%$ in flux measurement when using forced measurement when $\log_{10}(\text{S/N}) < 0.1$. This can be attributed to the injected centroid errors. No bias is present in the forced measurement of $\sigma_{xx}^2$ and $\sigma_{xy}^2$. Once again, the tighter error properties of forced measurement are clear.

Forced measurement is show in Figure 9. We find that the number of failures drops to 0 for all S/Ns. In the range $10 < \text{S/N} < 15$, the failure rate is 0%, and the percentage error in flux is $0.7\% \pm 1.2\%$. The percentage error in $\sigma_{xx}^2$ in this range is $0.13\% \pm 0.68\%$, whereas $\sigma_{xy}^2$ was recovered with an accuracy of $-0.1\% \pm 3.5\%$. The performance of this method becomes clear at lower S/N. In the range $0.1 < \text{S/N} < 1$, the failure rate is 0%, and the percentage error in flux is $-1.9\% \pm 21\%$. For cases in this range, $\sigma_{xx}^2$ was recovered with an accuracy of $0\% \pm 0.8\%$, while the percentage error in $\sigma_{xy}^2$ was $-0.2\% \pm 4.7\%$. This is much better than the previous case. We get an unbiased measurement with a tighter error spread.

Since forced measurement with a single iteration showed the best performance, we test this algorithm further. We select 3961 cases with a unique combination of flux, size, background, and ellipticity such that the S/N lies between 1 and 8. The ranges from which these values are selected are the same as before. We make 200 instances of a single source and measure each instance with forced single iteration using true guess values. We find the median of all 200 measurements and call this our best estimate from individual image measurements. These are shown as blue points in Figure 10. We also coadd the 200 instances and run the coadded image through the traditional moment matching algorithm. Because of the S/N range selected initially, convergence of the traditional moment matching algorithm on the coadd is guaranteed. These are shown as red dots on the plot. We can see that forced measurement performs significantly better. In the range of $\text{S/N} < 0.1$, the accuracy of forced measurement for flux is $0.18\% \pm 3.9\%$. The percentage error for $\sigma_{xx}^2$ is $-0.04\% \pm 0.8\%$, while for $\sigma_{xy}^2$, the percentage error is $-0.5\% \pm 4.9\%$. It has tighter error properties without bias.

In reality, however, sources are affected differently by atmospheric effects in each exposure. This causes random variations in their flux, centroids, and shape (Pier et al. 2003; Chang et al. 2012; Kerber et al. 2016). To simulate this, we follow the same method as before of simulating 200 cutouts for each combination of flux, size, background level, and ellipticity. However, in this case, we inject a 1 pixel error in the centroid and a 15% error in all three size variances ($\sigma_{xx}^2, \sigma_{yy}^2, \sigma_{xy}^2$) when simulating the source. This simulates the random size and centroid variation expected in multiple exposures. The errors are drawn from a random normal distribution. In addition to these errors, we also inject an error in our guess parameters. The error in guess flux and $\sigma^2$ is 15%. This effectively introduces an additional error in $\sigma^2$ and simulates the variation in flux expected in each cutout. The results are shown in Figure 11. We see that the flux calculation with the new method shows a few percent bias, which was determined to come from the centroid errors introduced. For cases where $\log_{10}(\text{S/N}) < 0.1$, the percentage error in flux is $-2.9\% \pm 4.1\%$, while for $\sigma_{xx}^2$, the error is $0.1\% \pm 1.6\%$. The percentage error for $\sigma_{xy}^2$ is $0.15\% \pm 4.8\%$. In the same S/N range, using a coadd to determine the same parameters gives an accuracy of $0.7\% \pm 10\%$ for flux, while $\sigma_{xx}^2$ has an error of $12.5\% \pm 22.6\%$. The percentage error in $\sigma_{xy}^2$ is $0\% \pm 185\%$. The shape parameters obtained from forced measurement ($\sigma_{xx}^2$ and $\sigma_{xy}^2$) show no bias. All the quantities have a significantly tighter error compared to measurements using the traditional moment matching method on the coadd. In Figure 12, we show the centroid measurements for the same case. We find no presence of bias and tighter error properties when using forced measurement.

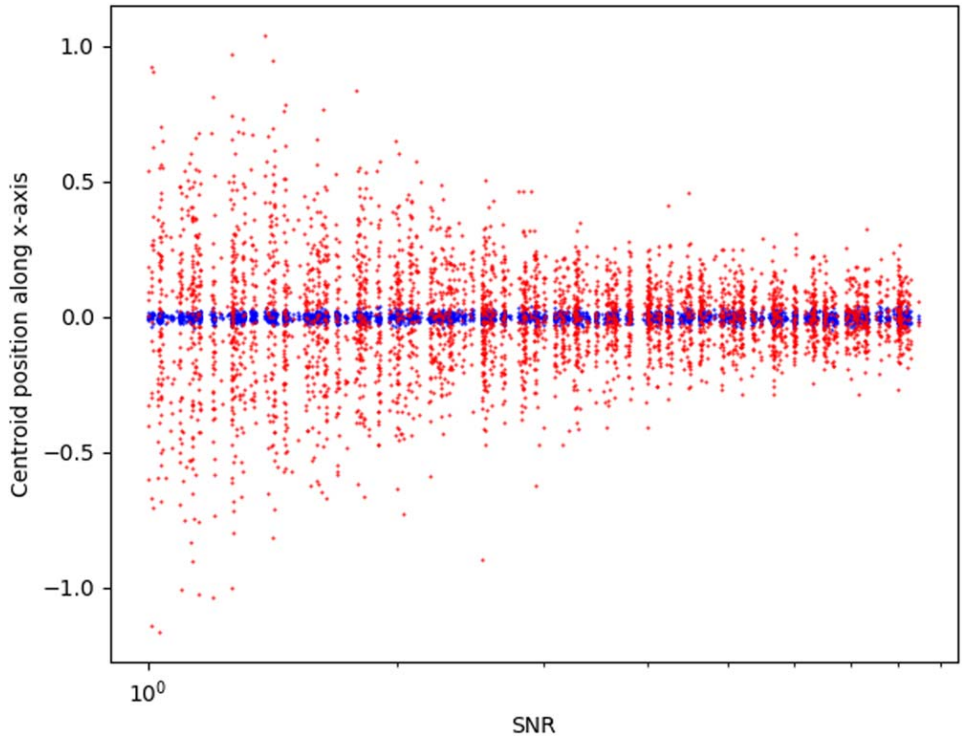

**Figure 12.** The red points show the *x*-position of the centroid from the coadd using the traditional moment matching algorithm, whereas the blue points show the median of the individual measurements obtained using forced measurement. The simulations are done under the same conditions as Figure 11.

Finally, we show the performance of this method in the presence of bias, since some level of bias in real use cases is to be expected. For this test, we give guess fluxes, $\sigma_{xx}^2$, $\sigma_{yy}^2$, and $\sigma_{xy}^2$, that are 10% larger than the true values. For each of the 200 cutouts simulated, we once again introduce a 1 pixel error in the centroid along each axis and 10% random errors in all three shape $\sigma$ to simulate the random variations expected in each cutout. We also inject 10% random errors in the guess flux and all three shape parameters. This method is exactly same as the previous case, except for the level of errors used and the additional bias component for the guess parameters. The results are shown in Figure 13. For the cases with $\log_{10}(S/N) < 0.1$, using the forced measurement method gives a $5\% \pm 4.3\%$ error in flux. The value of $\sigma_{xx}^2$ can be determined with an accuracy of $10.4\% \pm 1.3\%$, and $\sigma_{xy}^2$ can be determined with an accuracy of $10.5\% \pm 4.8\%$. In the same S/N range, using traditional moment matching on the coadd to determine the same parameters gives an accuracy of $1.7\% \pm 9.9\%$ for flux. The error in $\sigma_{xx}^2$ is $16.1\% \pm 26\%$, and the error in $\sigma_{xy}^2$ is $18.6\% \pm 177\%$. Flux values from the forced measurement are better than the initial biased guess while $\sigma_{xx}^2$ and $\sigma_{xy}^2$ are measured at the same level of input bias.

For $\sigma_{xx}^2$ and $\sigma_{xy}^2$ at higher S/N, forced measurement produces slightly worse results than the guess. However, this is not a huge issue since at higher S/N, the shape can be determined from the coadd with much better accuracy. This can then be used to construct a better guess value and improve the forced measurement. In the range $0.75 < \log_{10}(S/N) < 0.85$ using the proposed method, the percentage error in flux is $4.8\% \pm 1.8\%$. In this range, the error in $\sigma_{xx}^2$ is $11.9\% \pm 2.1\%$, while the error in $\sigma_{xy}^2$ is $10.6\% \pm 4.7\%$. In the same S/N range, using traditional moment matching on the coadd to determine the same parameters gives an accuracy of $0\% \pm 1.8\%$ when determining flux and an error of $12.1\% \pm 10\%$ when determining $\sigma_{xx}^2$. The percentage error for $\sigma_{xy}^2$ is $0.15\% \pm 29\%$.

### 4.2. Tests on WIYN-ODI Data

In this section, we show the characteristics of three different algorithms under different values of S/N: (1) the traditional moment matching algorithm, (2) forced measurement without truncation, and (3) forced measurement with truncation and bias correction. We note that flux obtained from forced measurement without truncation is identical to forced photometry. The data for this test have been obtained using the ODI instrument on the WIYN 3.5 m as a part of a weak lensing analysis program. The ODI consists of 30 CCDs arranged in a $5 \times 6$ pattern with a pixel scale of $0.''11$ pixel$^{-1}$. The seeing conditions during data acquisition varied between $0.''6$ and $3''$ with a median value of $1''$. All exposures are 60 s long and were obtained using one of the five wideband photometric filters available on ODI, namely, *u*, *g*, *r*, *i*, and *z*. The complete details of these data along with the weak lensing analysis will be published in a follow-up paper. We use SWarp (Bertin et al. 2002) for coadding 124 *i*-band and 144 *r*-band images. For coadding, it was found that using weight

$$W = 100 \frac{10^{Z-25}}{(S\sigma_b)^2} \tag{39}$$

is optimal, where $Z$ is the zero-point of the image, $S$ is the seeing, and $\sigma_b^2$ is the background variance. The coadded images were visually inspected, and several regions of artifacts were identified. These regions were masked from source detection and constitute less than 2% of the area of the coadded image. Coadding is also done individually for the *i* and *r* bands. Source detection is done using SExtractor (Bertin & Arnouts 1996) on the *i* + *r* coadded image with "DETECT_THRESH" and "ANALYSIS_THRESH" set to 0.8 and "DETECT_MINAREA" set to 5 pixels. After source detection, we measure the detected sources in the *i*-band coadd using the traditional moment matching algorithm. The process for making optimal cutouts in the coadd in order to reduce light contamination from nearby sources is nontrivial and is described in the follow-up paper. However, it was found for a vast majority of the sources that this does not make a significant difference. Using a square cutout based on the size measured from SExtractor is sufficient.

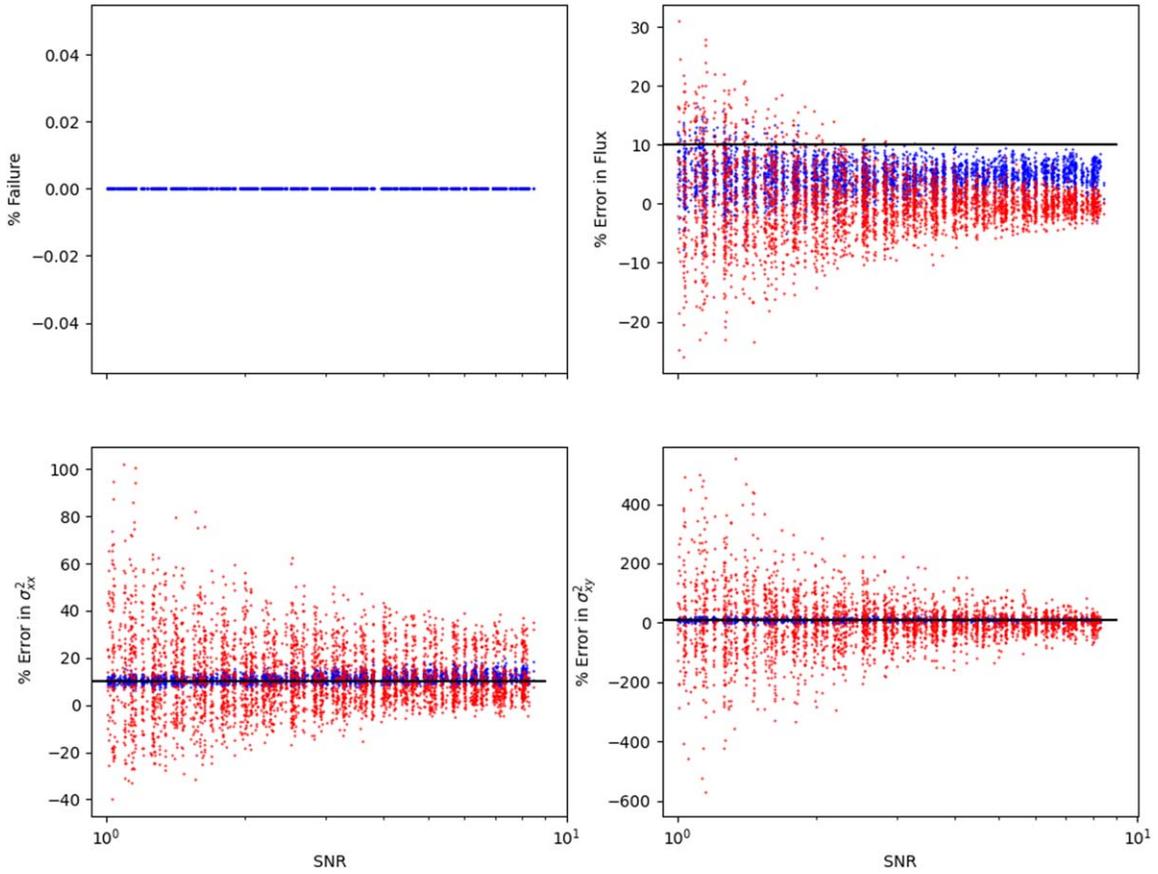

**Figure 13.** The red points show the coadd measurements using the traditional moment matching algorithm, whereas blue points show the median of the measurements obtained from the individual images using forced measurement. The *x*-axis is the S/N of individual images. We also inject an error of 10% in initial guess flux, $\sigma_{xx}^2$, $\sigma_{yy}^2$, and $\sigma_{xy}^2$. When simulating the individual sources, an error of 10% is injected in flux and $\sigma$, and an error of 1 pixel is introduced in the centroid. All guess parameters except the centroid have 10% overbias. This level is shown by the black line. In virtually all cases, forced measurement produces a better flux estimate. The accuracy at $\log_{10}(\text{S/N}) < 0.1$ using the forced measurement method is 5% ± 4.3% for flux; $\sigma_{xx}^2$ can be determined with an accuracy of 10.4% ± 1.3%, and $\sigma_{yy}^2$ can be determined with an accuracy of 10.5% ± 4.8%. In the same S/N range, using traditional moment matching on the coadd to determine the same parameter gives an accuracy of 1.7% ± 9.9% for flux, while the error for $\sigma_{xx}^2$ is 16.1% ± 26%. The percentage error for $\sigma_{xy}^2$ is 18.6% ± 177%.

The magnitude-versus-size graph of all sources as measured in the *i*-band coadd is shown in Figure 14. The vertical strip corresponds to stars that have a fixed size, i.e., the size of the PSF, but vary in magnitude. For this test, we select stars with *i*-band magnitude brightness in the range 17–22.5, while the range for size was from 2.8 pixels to 3.4 pixels. These limits were visually determined and are shown by the black bounding box in Figure 14. These sources will henceforth be called the test sample. It is clear from the plot that this sample is completely dominated by stars. All sources in Figure 14 were cross-matched with Gaia EDR3 (Brown et al. 2021) to determine a clean sample of stars. Sources that were successfully cross-matched are used for PSF interpolation and construction of guess values for forced measurement. We note that virtually all the cross-matched sources are a subset of the test sample.

Next, we measure sources in 146 *i*-band and 151 *r*-band images using the three different methods. First, all sources cross-matched with Gaia EDR3 and brighter than S/N 15 in a given frame were measured using the traditional moment matching algorithm. Next, we measure sources in the test sample. To measure a given source in the individual image, a 70 × 70 pixel cutout at the location of the source was created.

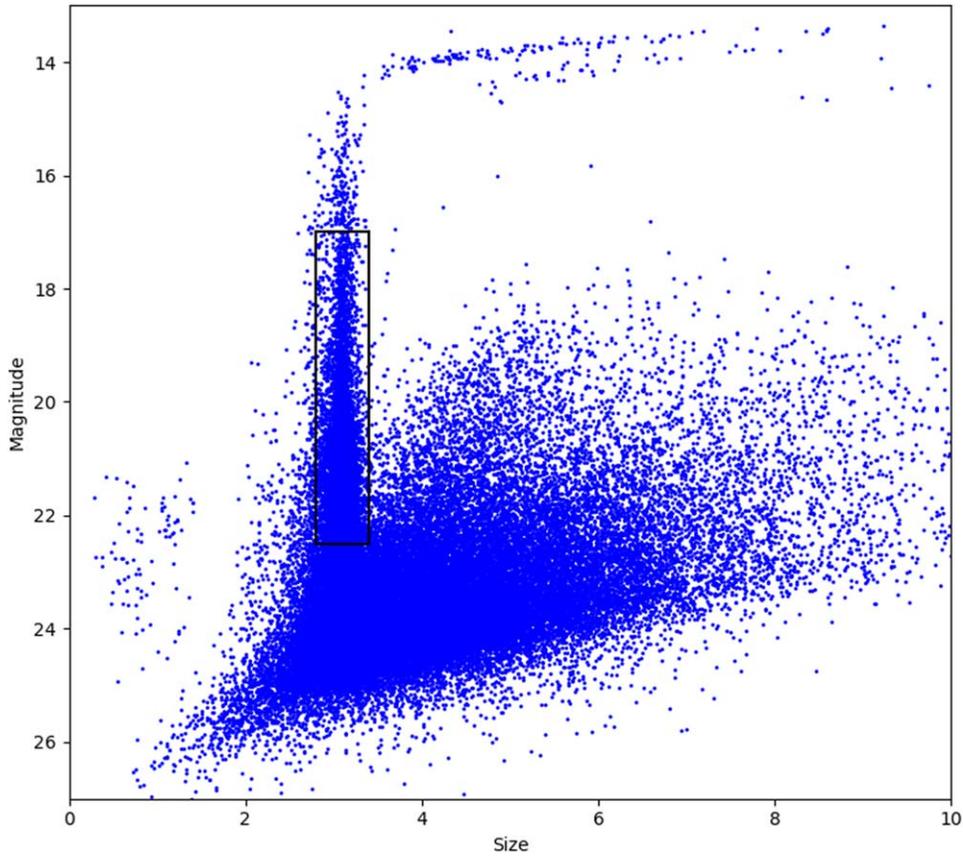

**Figure 14.** This plot shows the magnitude vs. size (in pixels) of all sources measured in the *i*-band coadd. The vertical strip corresponds to stars. Shown in the black box are all the sources for which individual frame measurements are performed, i.e., the test sample. The test sample includes sources with magnitude brightness in the range from 17th to 22.5th magnitude, while the size is limited from 2.8 to 3.4 pixels. These limits were determined visually, but as is clearly seen, almost all of these sources are likely to be stars.

If any negative or 0 values in the cutout are found, it is discarded. We also reject images with a PSF size greater than 7 pixels, which corresponds to a seeing size of 2.″4. To obtain the initial guess shape, we perform PSF interpolation using the 10 nearest stars with inverse distance as weight. This method of PSF interpolation was found to perform well (Gentile et al. 2012). The guess shape can be written as

$$\sigma_{xx}^2(\text{guess}) = \sum_{i=1}^{n} \sigma_{xx,i}^2 w(i), \qquad (40)$$

where $\sigma_{xx,i}^2$ is the measured value of the *i*th star and $w(i)$ is its corresponding weight. Weight $w(i)$ can be further simplified as

$$w(i) = \frac{1/d_i}{\sum_i 1/d_i}, \qquad (41)$$

where $d_i$ is the distance of the *i*th star from the source. Similarly, $\sigma_{yy}^2(\text{guess})$ and $\sigma_{xy}^2(\text{guess})$ were calculated. The interpolated PSF values were used as initial guesses for forced measurement, since the sources in the test sample are primarily stars.

The guess position of the stars in the individual images was determined from the coadd. Forced measurement also requires a guess flux when using truncation and bias correction. The guess flux was determined using a ratio of counts of the 10 nearest stars in the individual images and coadd. More specifically,

$$N_{\text{guess}} = N_{\text{coadd}} \times \left\langle \frac{N_{\text{image}}(\text{star})}{N_{\text{coadd}}(\text{star})} \right\rangle, \qquad (42)$$

where $N_{\text{guess}}$ is the guess flux and $N_{\text{coadd}}$ is the counts of the source in the coadd. $N_{\text{image}}(\text{star})/N_{\text{coadd}}(\text{star})$ is the ratio of counts in the image to the coadd of a nearby star that was used to interpolate the PSF. The angle bracket indicates the median ratio of the 10 nearest stars.

In order to compare the performance of the different algorithms, knowledge of the ground truth is crucial. This is only feasible in simulations. In real images, atmosphere, detector imperfections, and a multitude of other factors make it impossible for one to know the ground truth. Hence, we decided to use the guess parameters as the ground truth. The results are shown in Figure 15. Shown in blue is the traditional moment matching algorithm, red shows forced measurement without truncation, and black shows forced measurement with truncation and bias correction. As can be seen in the higher-S/N regime, the guess values and the values measured using the traditional moment matching algorithm are in excellent agreement. This shows that the guess values are a reasonably close approximation to the truth in most cases. The red and black points are slightly shifted along the *x*-axis for the sake of visual clarity. A measurement is considered to have failed if the flux is less than or equal to 0 or the variance of a source along the *x*- or *y*-axis is less than 0. Clearly, these outcomes are unphysical. Measurements where the measured size is greater than 6 times the size of the cutout are also rejected. In total, we have 565,132 valid measurements. The $x$-axis was divided into nine uniformly spaced bins in log space. The lowest S/N bin has 4280 samples, the lowest sample size of any bin. The second-highest S/N bin has 130,726 samples, the highest sample number of all bins. The median and standard deviation after $k = 3$ sigma clipping is plotted for each of the three methods in each bin. The performance of forced measurement with truncation and bias correction is clearly superior. At S/N $\sim 0.4$, when using forced measurement with truncation, the fractional error in flux is $0.2 \pm 1.0$, whereas the fractional error in $\sigma_{xx}^2$ is $-0.0 \pm 0.08$. The failure rate is 28%. Using forced measurement without truncation in this S/N range yields a failure rate of 82% and a fractional flux error of $2.8 \pm 1.8$. This shows the importance of truncation at low S/N. The traditional moment matching algorithm in this S/N range performs worst, with a failure rate of 91% and an average fractional flux error of $3.8 \pm 2.7$.

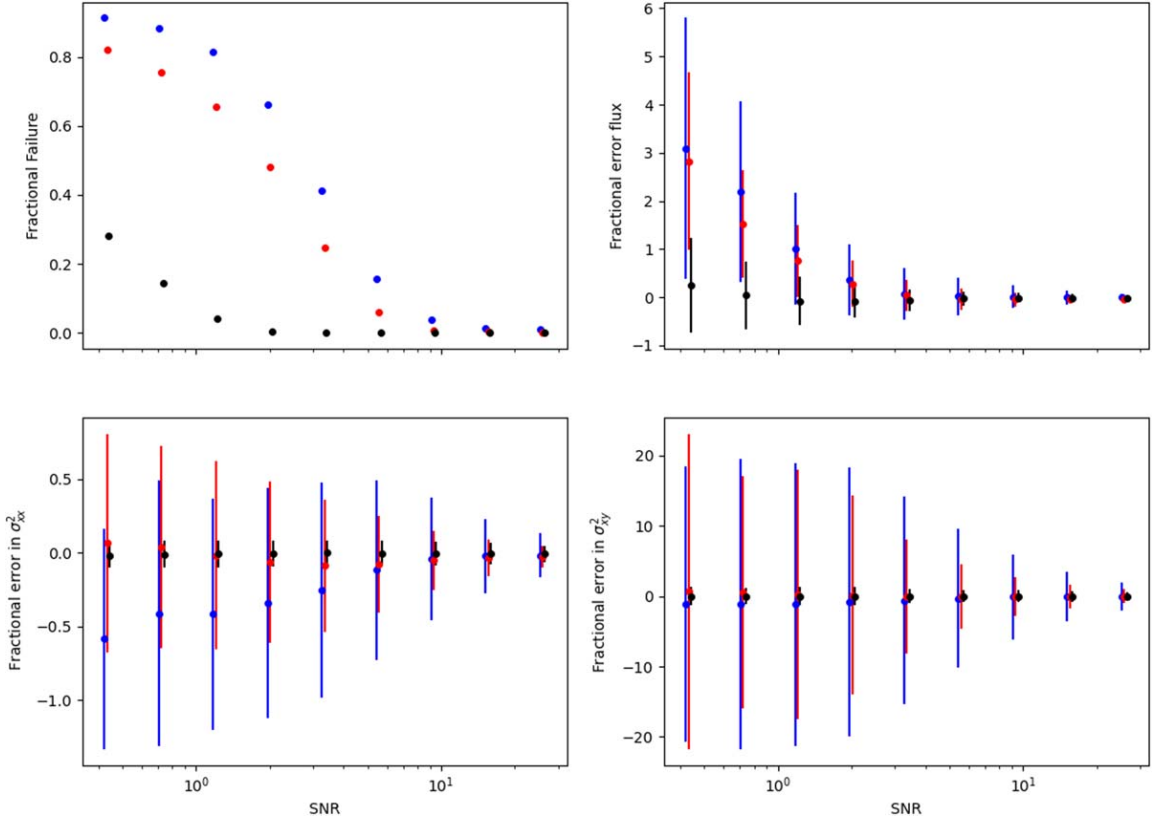

**Figure 15.** This plot shows the performance of the traditional moment matching algorithm in blue, forced measurement without truncation in red, and forced measurement with truncation and bias correction in black. Individual frame measurements were performed in the 146 $i$-band and 151 $r$-band images for all sources inside the black box shown in Figure 14. The $x$-axis is divided into nine equal bins in log space. The points and the error bars shown are the median and standard deviation after sigma clipping in each bin. It is clear that our method of forced measurement with bias and truncation correction outperforms in every aspect. The red and black data points have been shifted slightly on the $x$-axis for visual clarity.

## 5. Conclusions

In this paper, we have presented the intrinsic and background error components of flux, centroid, size, and ellipticity. Next, we show that the traditional moment matching algorithm fails at S/N $\lesssim 10$. To solve this, we propose a modification to the existing moment matching algorithm to measure the flux and shapes of sources at an arbitrarily low S/N. To do this, we apply the truncation (Equation (28)) to the background-subtracted image and run the forced measurement algorithm on this. We find that the flux and size parameters are overestimated. We correct for the overestimations using the function $f_1$ and $g_1$ (Equations (29) and (30)) to recover the original parameters. The failure rate of the new method drops to 0 in the range $0.1 < S/N < 1$ compared to a failure rate of $89.5\% \pm 3.6\%$ when using the traditional moment matching algorithm. In the same range of percentage error in flux, $\sigma_{xx}^2$ and $\sigma_{xy}^2$ reduce from an average of 396%, $-1.9\%$, and $-40\%$ to values that are consistent with 0. This method also worked when we allow for convergence, although convergence happens at extremely high S/N $\gtrsim 100$. This is not useful; hence, we do not consider it further. Finally, we show that we can use the forced measurement algorithm up to an extremely low S/N even in the presence of a substantial noise of $\sim$15% in $\sigma^2$, flux, and at least a 1 pixel error along each axis. Finally, we show that in the presence of 10% overbias in guess shape and flux along with 10% noise, we are still able to recover the original parameters satisfactorily. In the range $\log_{10}(S/N) < 0.1$ using forced measurement, the percentage error in flux is $5\% \pm 4.3\%$. The percentage error in $\sigma_{xx}^2$ is $10.4\% \pm 1.3\%$, while in the same range, the error in $\sigma_{xy}^2$ is $10.5\% \pm 4.8\%$. In the same range using the traditional moment matching algorithm on the coadd, we get errors of $1.7\% \pm 9.9\%$ in flux. The percentage error for $\sigma_{xx}^2$ is $16.1\% \pm 26\%$, whereas the error for $\sigma_{xy}^2$ is $18.6\% \pm 177\%$. While the level of errors when using forced and traditional moment matching are comparable, we note that in spite of a biased guess value, using forced measurement gives us a better estimate of flux than our initial guess and $\sigma^2$ values that are not worse than the biased initial guess. We also get significantly tighter error properties. On applying this method to measure point sources in data obtained from WIYN-ODI, at S/N $\sim$ 0.4 we get a failure rate of 28%, and the fractional error in flux is $0.2 \pm 1.0$. At the same S/N, forced photometry produces a fractional error in flux of $2.8 \pm 1.8$ with a failure rate of 82%. Comparing the failure rates, we find that the proposed method allows us to measure sources having a 7 times lower S/N than is possible with traditional methods. In conclusion, this a method to produce better measurements in the presence of reasonable estimates in most cases. In extreme cases with a significant level of bias and error, it produces some measurements that are no worse than the original guess. Future studies with real images will refine the techniques.

## Acknowledgments

We thank the anonymous referee for helpful comments. The authors would like to thank Purdue University for continued support. We also thank the Purdue Rosen Center for Advanced Computing (RCAC) for access to computing facilities that have been extensively used in this paper. This data analysis has been done on python, and the authors acknowledge the use of astropy, numpy, scipy, and matplotlib.

## ORCID iDs

A. Dutta  https://orcid.org/0009-0000-1088-4653
J. R. Peterson  https://orcid.org/0000-0001-5471-9609